\pdfoutput=1

\documentclass[a4paper]{article}

\usepackage{icrc2013}
\usepackage{todonotes}
\usepackage{textcomp}

\title{The Tunka-Rex antenna station}
\newcommand{\etal}{\MakeLowercase{\textit{et al. }}} % "et al."
\shorttitle{R.~Hiller \etal Tunka-Rex station}

\authors{R.~Hiller$^{1}$, N.M.~Budnev$^{2}$, O.A.~Gress$^{2}$, A.~Haungs$^{1}$, T.~Huege$^{1}$, Y.~Kazarina$^{2}$, M.~Kleifges$^{3}$, A.~Konstantinov$^{4}$, E.N.~Konstantinov$^{2}$, E.E.~Korosteleva$^{4}$, D.~Kostunin$^{1}$, O.~Kr\"omer$^{3}$, L.A.~Kuzmichev$^{4}$, R.R.~Mirgazov$^{2}$, L.~Pankov$^{2}$, V.V.~Prosin$^{4}$, G.I.~Rubtsov$^{5}$, C.~R\"uhle$^{3}$, F.G.~Schr\"oder$^{1}$, E.~Svetnitsky$^{2}$, R.~Wischnewski$^{6}$, A.~Zagorodnikov$^{2}$ (Tunka-Rex Collaboration)}

\afiliations{$^{1}$Institut f\"ur Kernphysik, Karlsruhe Institute of Technology (KIT), Germany\\ $^{2}$Institute of Applied Physics ISU, Irkutsk, Russia\\ $^{3}$Institut f\"ur Prozessdatenverarbeitung und Elektronik, Karlsruhe Institute of Technology (KIT), Germany\\ $^{4}$Skobeltsyn Institute of Nuclear Physics MSU, Moscow, Russia\\ $^{5}$ Institute for Nuclear Research of the Russian Academy of Sciences, Moscow, Russia\\ $^{6}$DESY, Zeuthen, Germany}

\email{roman.hiller@kit.edu}

\abstract{Tunka-Rex is the radio extension of Tunka-133, a 1~km$^{2}$ air-Cherenkov Detector for air showers in Siberia. Tunka-Rex began operation on October 8th 2012 with 20 radio antennas.
Its main goals are to explore the possible precision of the radio detection technique in determination of primary energy and mass.
Each radio antenna station consists of two perpendicular aligned active SALLA antennas, which receive the radio signal from air showers. The preamplified radio signal is transmitted to local cluster centers of the Tunka-133 DAQ, where it is filtered, amplified and digitized.
To reconstruct the radio signal it is crucial to understand how it is affected in each of these steps. Thus, we have studied the combined response of the antenna, with its directional pattern and the analog electronics chain, consisting of a Low-Noise Amplifier and a filter amplifier. 
We discuss the hardware setup of Tunka-Rex and how a description of its response is obtained. Furthermore, we estimate systematic uncertainties on the reconstructed radio signal due to hardware effects (e.g., slight variations of the electronics properties). Finally, we present background measurements with the actual Tunka-Rex antennas.}

\begin{document}
\maketitle

%Begin the section.
\section{Introduction}
One candidate for a new, cost effective detector type for cosmic ray air showers are radio detectors. Radio emission from air showers was discovered more than 40 years ago \cite{radio1971}. After some initial effort it became a very active research field once again in recent years due to the methods of modern signal processing. Experiments like LOPES \cite{FalckeNature2005} have already proven the principal feasibility of such a detector. Now its capabilities and especially its precision in determination of energy and shower maximum have to be investigated further to test its applicability as stand alone detector or part of a hybrid detector.\\
With this goal, the Tunka Radio Extension (Tunka-Rex)\cite{TunkaRexICRC2013} was deployed at the Tunka-133 \cite{TunkaICRC2013} site. Tunka-133 is a non-imaging air-Cherenkov detector, consisting of 133 upward looking PMTs, organized in hexagonal clusters of 7. They are distributed over 1~km$^{2}$. The Tunka-Rex detector currently consists of 20 stations, one next to each cluster center of Tunka-133, resulting in a spacing of about 200~m between the antennas. It uses the trigger and DAQ system from Tunka-133. It started operation in October 2012 and already successfully finished the first measurement season.\\
In this paper, we first describe the basic properties of the detector. Followed by an overview of how the hardware response was obtained and how the initial radio signal is reconstructed from it. Then we estimate systematic uncertainties on the reconstructed signal amplitude. Finally we discuss the different kinds of background.    
\section{Hardware Properties of Tunka-Rex} 
Each station consists of two antennas aligned perpendicularly, sensitive to two linear independent components of the electrical field. Thus, with additional information about the incoming direction of the electric field, it is possible to reconstruct all three components of the electrical field. The direction is obtained from combined information of signal arrival time of different stations.\\ 
The components of each channel  are an antenna, a Low Noise Amplifier (LNA), cables, a filter and ADCs.
\begin{figure}[tb]
\includegraphics[width=0.46\textwidth]{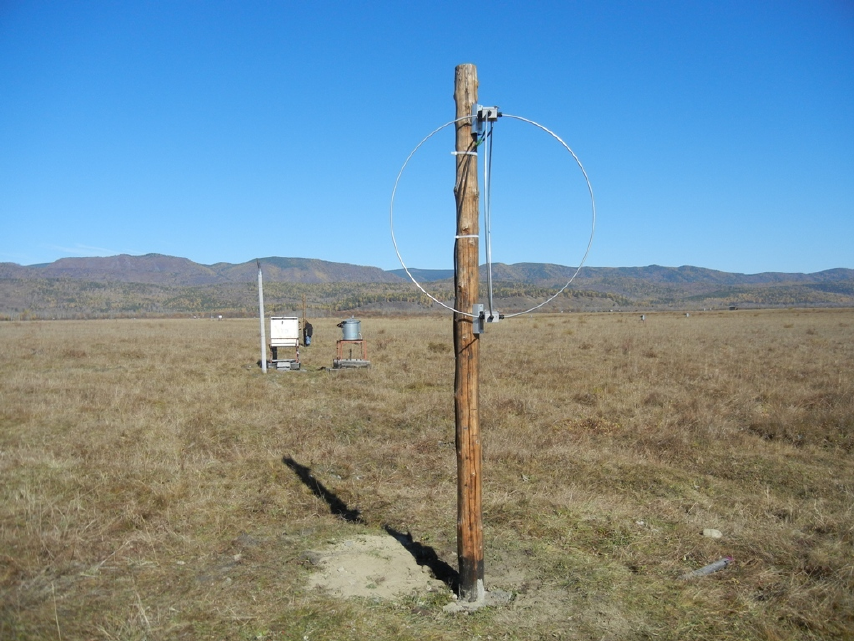}
\label{fig:SALLA}
\caption{A Tunka-Rex antenna in front of a Tunka-133 cluster center box and its central PMT.}
\end{figure}
The antenna used for Tunka-Rex is the SALLA \cite{KroemerICRC2009}\cite{AERAantennaPaper2012} with 120~cm diameter. It is a relatively economic and rugged type of Beverage antenna. With a half-power beam width of 140$^{\circ}$ it has a good sky coverage up to zenith angles of 70$^\circ$. That is well above the analysis cutoff of Tunka-133 of 50$^{\circ}$. It has a load pointed towards the ground. Thus reduces its gain from below and this it has little dependency on ground conditions. The downside of the SALLA design is its relative low gain.
Unlike most radio experiments the antennas in Tunka-Rex are not aligned along the north-south and east-west axis, but rotated by 45$^{\circ}$, like in LOFAR \cite{ARENALOFAR}. Since the radio signal from cosmic ray air showers is predominantly east-west polarized \cite{polarization}, this should result in more antennas with signal in both channels but also less events with signal in at least one channel. Thus it is a trade off between efficiency and reconstruction quality close to the threshold.\\
The SALLA we use in Tunka-Rex is an enhanced version of the one developed as candidate for LOPES\cite{FalckeNature2005} and later AERA\cite{MelissasAERA_ARENA2012}. It is an active antenna, i.e. it has a Low Noise Amplifier integrated in the antenna in the upper box, which is connecting the arcs. Main part of the LNA is the commercial amplifier IC MGA-62563 with very good linearity and amplification of about 24~dB.\\
30~m of RG213 cable connect the LNA to the filter inside the cluster center box and attenuate the signal by 1~dB. The interior of the cluster center boxes is heated so the filter and the ADCs operate under controlled temperature. The filter-amplifier is the final part of the analogue chain. It filters frequencies outside the design frequency band of 30-80~MHz and amplifies the remaining signal by another 32~dB.\\
Finally the signal is digitized by the commercially available AD9430 at a sampling rate of 200~MHz with a depth of 12 bit, thus we operate in the first Nyquist domain for the design frequency band. Each recorded trace is $5~\mathrm{ns}\cdot 1024 \approx 5$~\textmu s long. It is centered around the trigger time. Since the trigger comes from the Tunka-133 PMTs with their 50~m longer cables, we expect the antenna signal before the trigger time.

\begin{figure}[]
\includegraphics[width=0.5\textwidth]{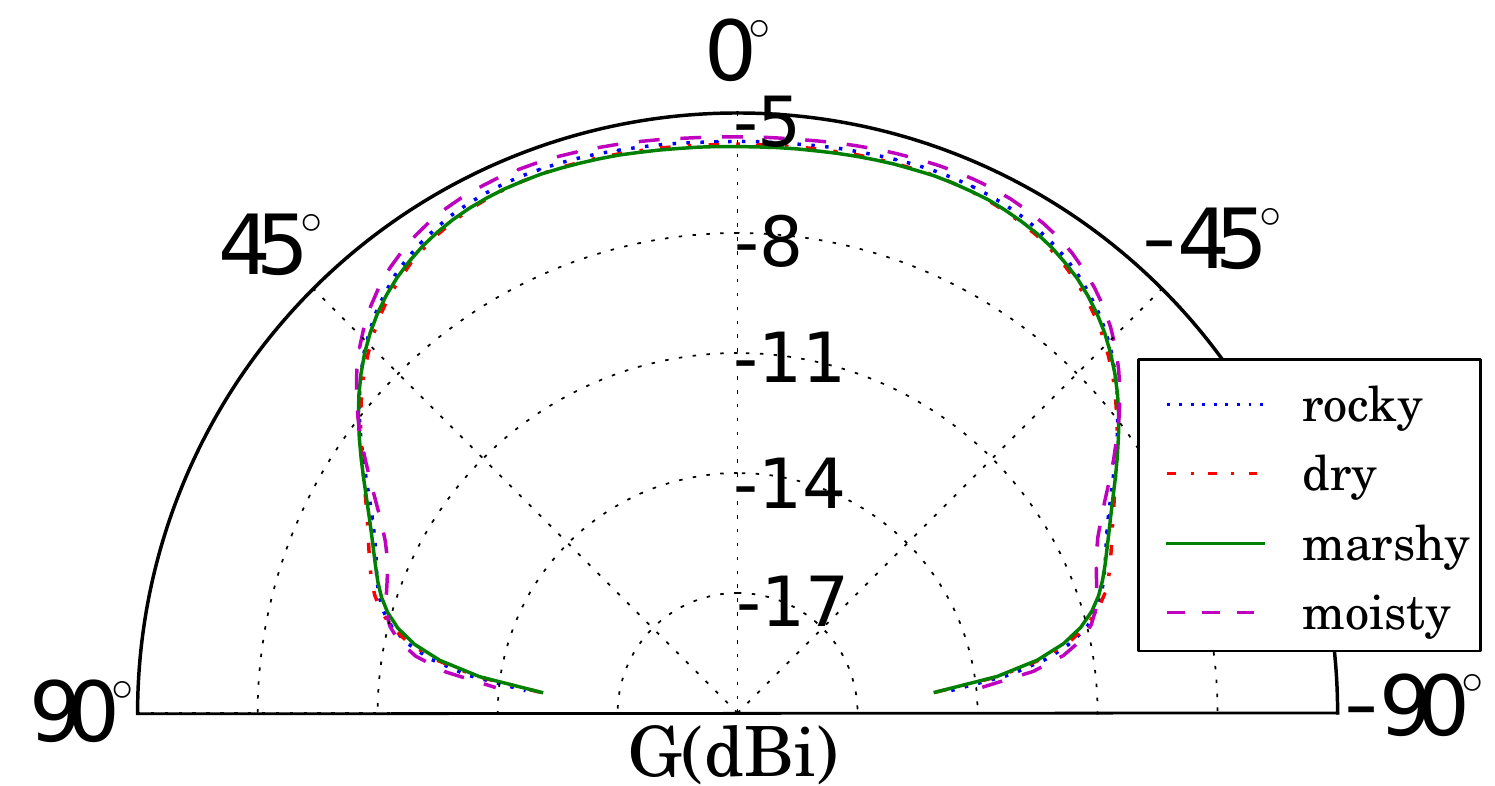}
\label{fig:SALLA_ground}
\caption{Gain of the SALLA in the vertical plane for a frequency of 50 MHz. The SALLA is aligned parallel to the shown plane. The different lines correspond to different ground types.}
\end{figure}
\section{Hardware response and E-field reconstruction}
\begin{figure}[]
\includegraphics[width=0.48\textwidth]{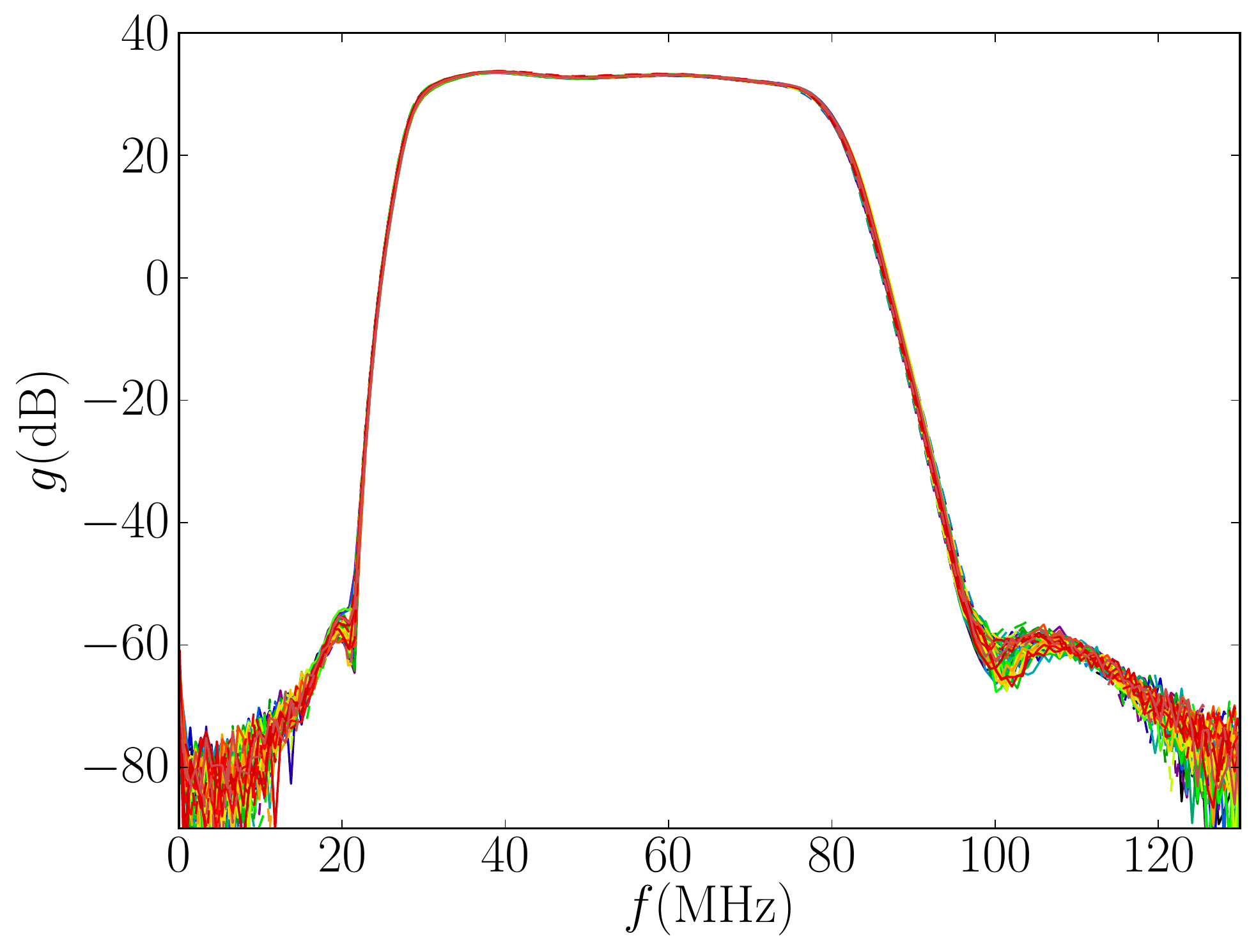}
\label{fig:filter_gain}
\caption{Power gain of the filter amplifiers. Different lines correspond to different filters. The filters amplification varies about 0.3~dB in the 30-80~MHz range from filter to filter.}
\end{figure}

The radio signal can be described by its electrical field strength $\vec E$. To retrieve it from the recorded ADC counts one has to understand how the signal is affected by all the hardware components .\\
\begin{figure}[tb]
\begin{center}
\includegraphics[width=0.35\textwidth]{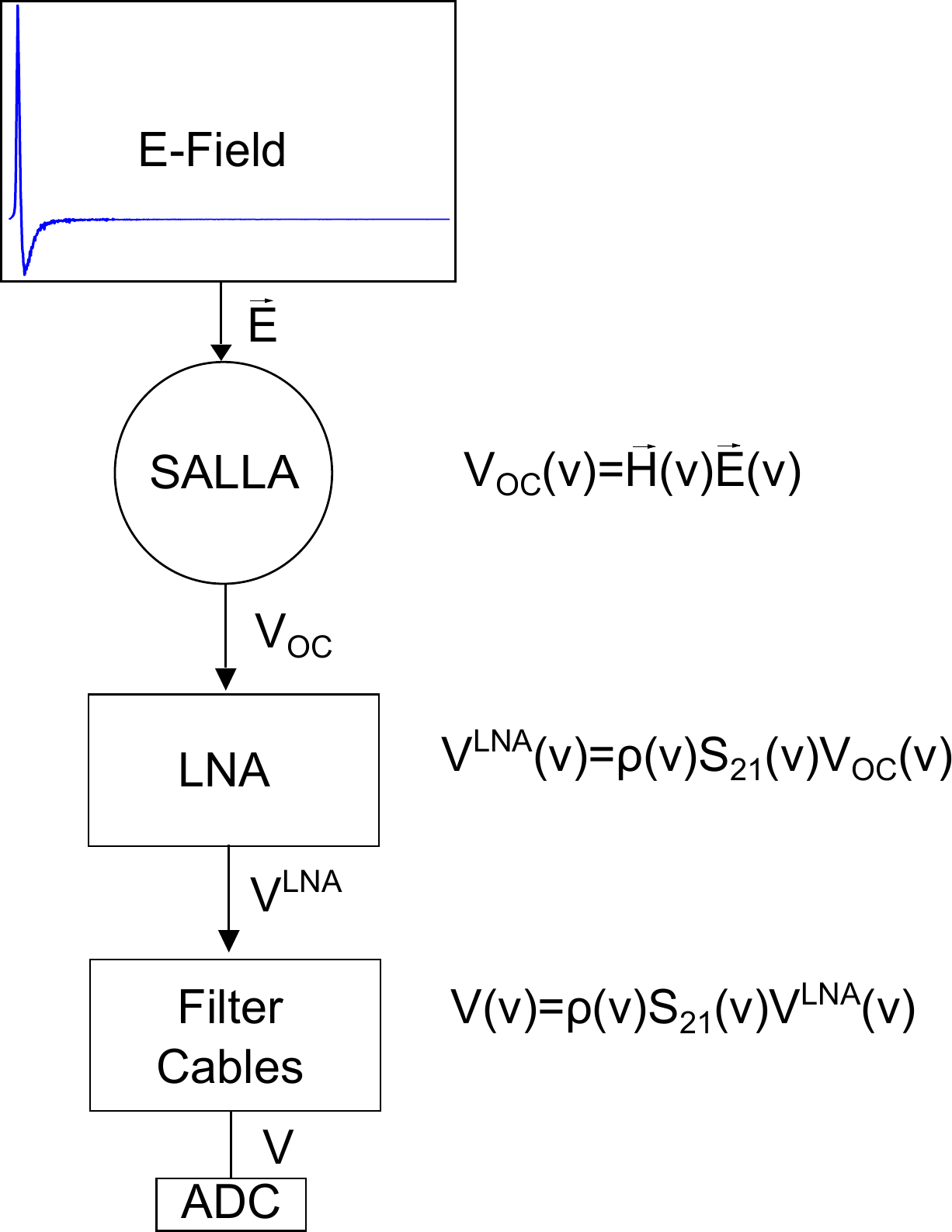}
\end{center}
\label{fig:SALLA}
\caption{Signal chain of a Tunka-Rex antenna with the corresponding transformations.}
\end{figure}
The first stage is the antenna. It receives the radio signal and converts it to an open circuit voltage at its foot point $V_{OC}$. In frequency space this can simply be described as a product of E-field and vector effective length H of the antenna
\begin{equation}
\label{eq:antenna_trafo}
\vec{U}(\nu) = H(\nu) \cdot \vec E(\nu) .
\end{equation}
To invert eq. \ref{eq:antenna_trafo} one only needs to know the vector effective length. One possibility to access it is to calculate it, e.g. with an antenna simulation program. We used the NEC2 software package \cite{nec2} to calculate the far field strength of the antenna under well defined conditions. H can then be obtained with the reciprocity theorem.
Another option is to retrieve H from an antenna calibration. We chose a composite approach: take the directivity pattern from simulations and check it with a calibrated source. For the amplitude calibration we used the same equipment and method as for LOPES \cite{LOPESCalibration}.\\
From the antenna terminal the signal is transferred to the analog chain. This transfer is not trivial, since the antenna impedance varies a lot over the bandwidth and the mismatch to the LNA input is a major influence. It can be calculated using an Thevenin equivalent diagram of the series connected SALLA impedance and the LNA input impedance. The Voltage drop over the LNA is only a fraction of $V_{OC}$, described by the matching factor \cite{AERAantennaPaper2012}
\begin{equation}
\label{eq:matching}
\rho = \frac{Z_{LNA}}{Z_{SALLA}+Z_{LNA}}.
\end{equation}
We took the SALLA impedance from the NEC2 simulations. To get the LNA input impedance a network analyzer was used to measure scattering parameters of the LNA. The input impedance of the LNA can be obtained from the input reflection $S_{11}$
\begin{equation}
\label{eq:reflection_factor}  
Z_{LNA}=Z_{NWA}\cdot\frac{1+S_{11}}{1-S_{11}}.
\end{equation}
Next the signal has to traverse the analog chain. As for the antenna this is described by convolving the input signal with the response function of the analog chain. In frequency space this is
\begin{equation}
\label{eq:response}
V_{out}(\nu) = V_{in}(\nu)\cdot S_{21}(\nu).
\end{equation}
The response functions of  LNAs, cables and filters were measured separately with a network analyzer. In fig. \ref{fig:filter_gain} the power gain for the measured amplitudes for forward transmission of the filter amplifier are shown.\\
The parts within the analog chain, as well as the network analyzer ports, are matched to 50~$\mathrm{\Omega}$. Therefore the forward transmission $S_{21}$ of cable and filter as their response function, measured with the network analyzer, already takes their input reflection into account.\\ 
For the LNA the situation is different, since it has the antenna attached to its input and therefore does not operate in a 50~$\mathrm{\Omega}$ system. With the matching factor we already obtained the fraction of signal transferred to the LNA. To get the factor by which the transferred signal is amplified we have to take reflection in the 50~$\Omega$ system into account by normalizing the measured $S_{21}$ according to the input reflection $S_{11}$
\begin{equation}
\label{eq:S21_norm}
S_{21}’=\frac{S_{21}^{NWA}}{1+\left|S_{11}^{NWA}\right|}
\end{equation}
Backward transmission $S_{12}$ and output reflection $S_{22}$ are not taken into account, because they are not expected to have any impact on the signal reconstruction.
\section{Systematic uncertainties on the amplitude reconstruction}
\begin{figure}[]
\includegraphics[width=0.5\textwidth]{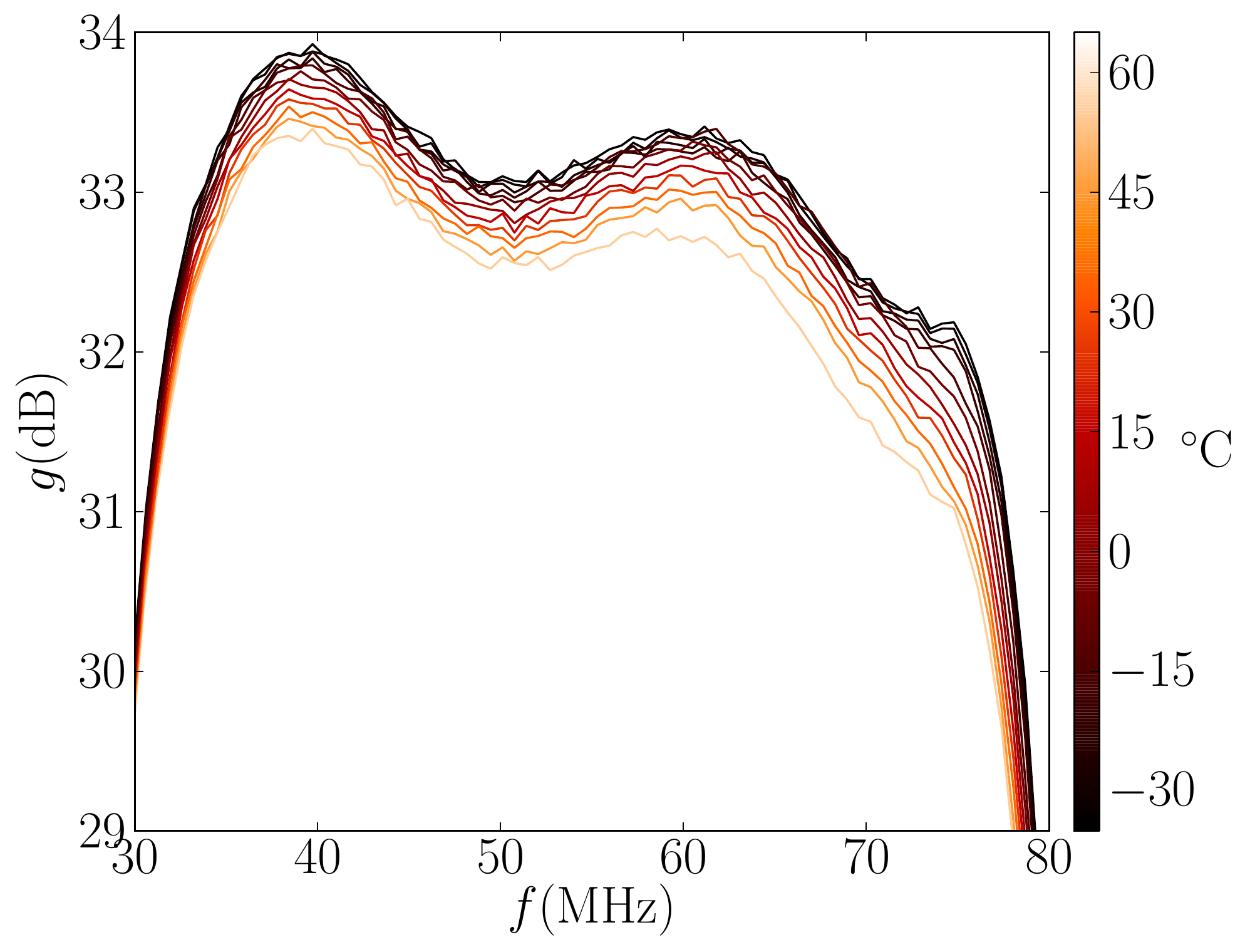}
\label{fig:filter_gain_temp}
\caption{Power gain of a filter amplifier at different temperatures. The filter was cooled or heated in a temperature chamber to -35~$^\circ$C and up to 65~$^\circ$C. }
\end{figure}

In the simulation as well as in the calibration we obtain the response of a single antenna. Of course not all antennas are exactly the same. Their parameters vary slightly from antenna to antenna and so does their response. Using the response from one antenna to reconstruct the E-field received by other antennas introduces a systematic uncertainty. To estimate this uncertainty we repeated the antenna simulation for different parameters. Then, we take a set of radio signals from REAS~3.11 simulations \cite{REAS}, convolve them with the modified antennas response and compare the resulting signal amplitude with the one from the standard response. As an estimate for the uncertainty we take the standard deviation of the residuals.\\
We simulated 4 proton showers each with zenith angles 0-45$^{\circ}$ in 15$^{\circ}$ steps, azimuth 0-90$^{\circ}$ in 30$^{\circ}$ steps, energy 10$^{17}$~eV and evaluated the radio signal at the antenna positions 500~m or closer to the core.\\
The uncertainties can be divided in two categories. The first is a relative amplitude uncertainty, changing from antenna to antenna in random directions. The second is an uncertainty, pointing in the same direction for all antennas. The following values were found for the relative amplitude uncertainty of each antenna:
\begin{itemize}
\item 1.3\% due to an varying antenna height up to 3~cm
\item 1.0\% due to an azimuth alignment uncertainty of 1$^{\circ}$
\item 0.4\% due to zenith alignment uncertainty of 1$^{\circ}$
\item 1.5\% due to deviations in the arc radius of the SALLA of 1~cm
\item 1.4\% due to production fluctuations in the load
\end{itemize}
Adding up to a total of 2.7\%, assuming that the individual uncertainties are uncorrelated.\\
Another uncertainty comes from the ground conditions. 5 different grounds have been tested, ranging from dry to marshy . The ground is expected to be approximately the same for all antennas, but remains unmonitored. Therefore another 1.0\% uncertainty has to be added, highly correlated for all antennas.\\
%The transmission paramaters are only measured with a precision of about ??. 
Further uncertainties arise from the temperature dependence of the analog chain response. It was investigated by repeating the transmission parameter measurements at different temperatures. Therefore, each part was heated or cooled to temperatures from -35$^{\circ}$~C to 65$^{\circ}$~C in a temperature chamber during the measurement. In fig. \ref{fig:filter_gain_temp} the amplitude of $S_{21}$ measurement is shown at different temperatures. Since at Tunka-Rex the filter is located in the cluster center box which is heated, its temperature will usually be around 20$^{\circ}$~C and won't significantly contribute to the uncertainty due to temperature. The LNA on the other hand is located next to the antenna, outside the box. Therefore its temperature will usually be below 0$^{\circ}$~C during winter and can drop down to -30$^{\circ}$~C regularly. Since the response at 20$^{\circ}$~C is used at the moment, an uncertainty of 4\% on the absolute amplitude scale has to be added.

\section{Background at Tunka-Rex}
\begin{figure}[t]
\includegraphics[angle=-90,width=0.5\textwidth]{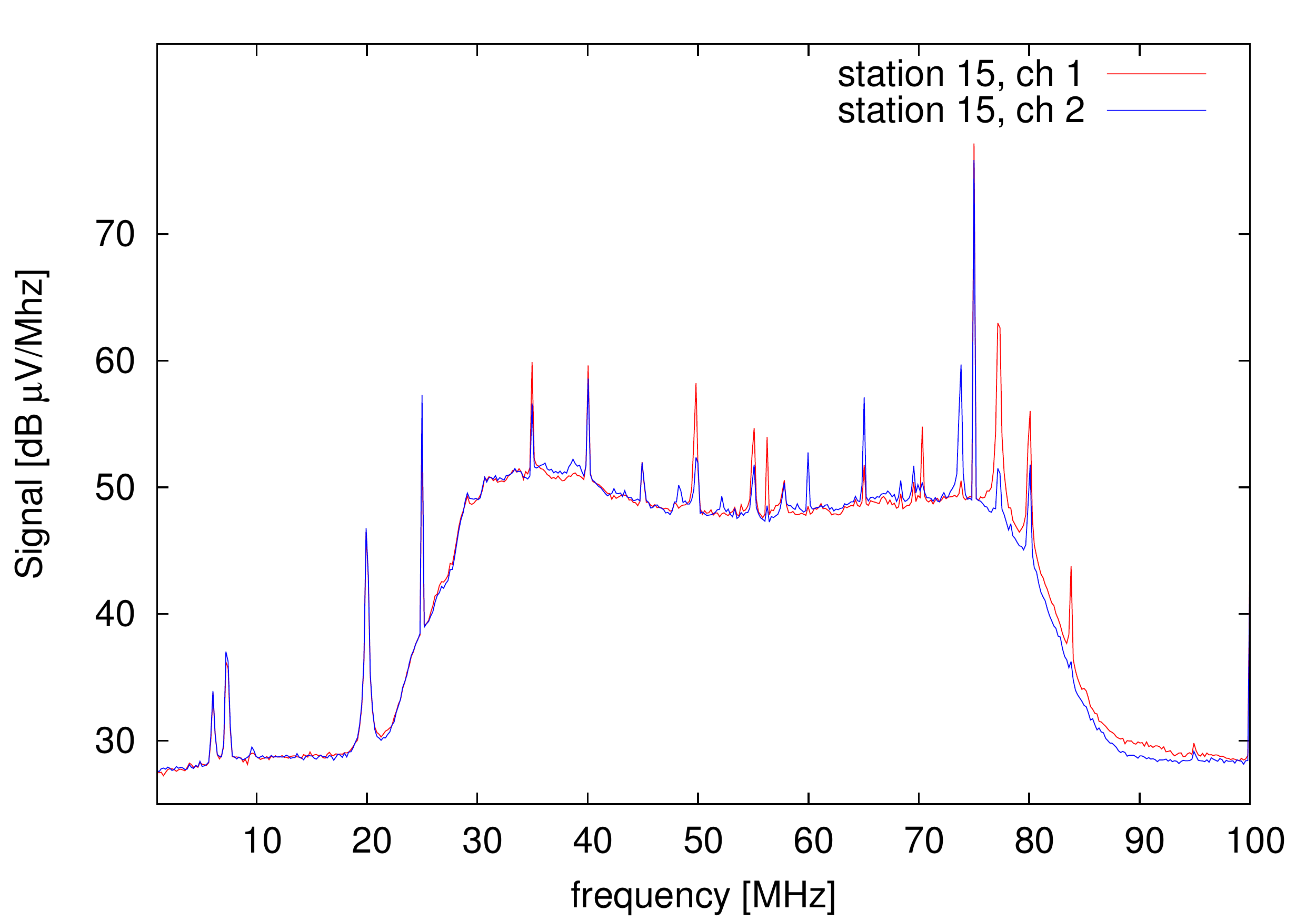}
\label{fig:spectrum}
\caption{The mean background spectrum in one night of operation, as measured with Tunka-Rex station 15.}
\end{figure}

To understand the measured signal and to optimize the signal-to-noise ratio, it is important to study the radio background, as it is the main uncertainty in the signal reconstruction.
There are three kinds of background sources: galactic background, thermal noise and human made background.\\
Galactic background is a broad band noise with a temperature of several 1000~K in the 30-80 MHz band. Thermal noise originates mainly from the SALLA load and the LNA. Its power is estimated to be in the same order of magnitude as the galactic background.\\
Galactic background and thermal noise are unavoidable broad band sources, which a radio signal has to surpass in order to be distinguishable from them. Human made background on the other hand is often narrow in frequency or time and can therefore be reduced by filtering. With this purpose the analog filter was included, to avoid strong signals from commercial radio below 30 and above 80 MHz. Looking in the spectrum of a Tunka-Rex antenna in fig. \ref{fig:spectrum} one can identify several narrow band radio sources, which are partially understood: 
\begin{itemize}
\item the 20~MHz line is from the heating controller
\item the lines at 25, 50 and 75 MHz are from other cluster box electronics, since they are also seen by the PMTs of Tunka-133
\end{itemize}
These narrow band lines can be filtered digitally and thus are no problem for our analysis.
Other sources within our band are not yet understood and there are ongoing analysis about their origin:
\begin{itemize}
\item there are several narrow lines at multiples of 5~MHz
\item from time to time there is a broad noise bump at 30-40~MHz
\end{itemize}
To investigate further what the origins of these sources are and also for the calibration, one antenna station was build at the KIT in Karlsruhe, Germany. Thus, changing the DAQ and location of the antenna. A first simple comparison of traces from the Tunka site and Karlsruhe shows that traces from Karlsruhe show a lot more narrow peaks, typical for densely populated city regions. The underlying broad band noise seems to be comparable. In future we plan to compare also the time dependence of the background.
%\begin{figure}[h!]
%\includegraphics[width=0.5\textwidth]{trace_all.png}
%\label{fig:traces}
%\caption{Traces of a Tunka-Rex station on the Tunka site and at KIT, Germany. }
%\end{figure}
\section{Conclusion}
A hardware description for the Tunka-Rex antenna was obtained by combining response of the passive antenna structure from simulations with the measured response of amplifier, cables and filter. It will be checked with a calibrated reference source. By simulating variations of the antenna, a systematic uncertainty of 2.7\% was derived as relative uncertainty between amplitudes of different antennas. Furthermore, an uncertainty of 5\% was found for the uncertainty of the absolute amplitude scale due to unmonitored ground conditions and temperature.  This can be tolerated when aiming for an energy precision of 10\%. While some sources of background at the Tunka-Rex site are already found, further analyses are currently conducted to identify the remaining ones. 

\vspace*{0.5cm}
\footnotesize{{\bf Acknowledgment:}{We acknowledge the support of the Russian Federation Ministry of Education and Science (G/C16.518.11.7051, 14.740.11.0890, 16.518.11.7051, P681, 14.B37.21.0785, N 14.518.11.7046), the Russian Foundation for Basic research (grants 10-02-00222, 11-02-12138, 12-02-10001, 12-02-91323), the President of the Russian Federation (grant MK-1170.2013.2), the Helmholtz association (grant HRJRG-303).}}

\end{document}